\documentclass[aip, jcp, reprint]{revtex4-2}

\usepackage{amsmath}
\usepackage{amssymb}
\usepackage{graphicx}
\usepackage{hyperref}
\usepackage[capitalize]{cleveref}
\crefname{equation}{}{}
\crefname{section}{section}{section}
\crefname{figure}{Fig.}{Fig.}
\crefname{table}{Table}{Table}
\renewcommand{\vec}[1]{\mathbf{#1}}
\newcommand{\WD}{\mathbf{D}}
\newcommand{\Wd}{\mathbf{d}}

\newcommand{\CC}{\mathrm{K}}
\newcommand{\nCC}{\hat{\CC}}
\newcommand{\OG}{{\text{O}(3)}}
\newcommand{\bfl}{\mathfrak{b}}
\DeclareMathOperator{\arctantwo}{arctan2}

\begin{document}

\title{Point Group Analysis in Particle Simulation Data}
\author{Michael Engel}
\affiliation{Institute for Multiscale Simulation, IZNF, Friedrich-Alexander-University Erlangen-N\"urnberg, Cauerstrasse 3, 91058 Erlangen, Germany}

\date{\today}

\begin{abstract}
A routine crystallography technique, crystal structure analysis, is rarely performed in computational condensed matter research.
The lack of methods to identify and characterize crystal structures reliably in particle simulation data complicates the comparison of simulation outcomes to experiment and the discovery of new materials.
Algorithms are sought that not only classify local structure but also analyze the type and degree of crystallographic order.
Here, we develop an algorithm that analyzes point group symmetry directly from particle coordinates.
The algorithm operates on functions defined on the surface of the sphere, such as the bond orientational order diagram.
Other use cases are the orientation of crystals and adoption as generalized order parameters for detecting the appearance of order as well as following its development.
\end{abstract}
\maketitle

%%%%%%%%%%%%%%%%%%%%%%%%%%%%%%%%%%%%%%%%%%%%%%%%%%%%%%%%%%%%%%%%%%%%%%%%%%%%%%%%%%%%%%%%%%%%%%%%%%%%
%%%%%%%%%%%%%%%%%%%%%%%%%%%%%%%%%%%%%%%%%%%%%%%%%%%%%%%%%%%%%%%%%%%%%%%%%%%%%%%%%%%%%%%%%%%%%%%%%%%%
%%%%%%%%%%%%%%%%%%%%%%%%%%%%%%%%%%%%%%%%%%%%%%%%%%%%%%%%%%%%%%%%%%%%%%%%%%%%%%%%%%%%%%%%%%%%%%%%%%%%

\section{Introduction}

The structure of even highly complex crystals and macromolecules can now be solved routinely~\cite{Massa2010}.
For this purpose, Bragg peak positions and intensities are measured via X-ray, neutron, or electron scattering.
This data is then input into a computer program to determine the space group and unit cell decoration by exploiting properties of the Fourier transform.
Despite success as an experimental technique, crystal structure analysis of simulation data is in its infancy.
This is surprising because simulation provides access to the trajectories of all particles and thus in principle much more data than any experiment.
However, a complicating factor is that system sizes and equilibration times in simulation are orders of magnitude smaller.
As a result, the resolution and quality of numerical diffraction patterns is usually too low for the application of traditional crystal structure solution techniques.
Instead, an alternative method operating directly on the particle coordinates is required.

Computer power and our understanding of ordering processes are advancing rapidly.
More and more simulations nucleate and grow crystals, even highly complex ones, directly from the melt~\cite{Walsh2009,Damasceno2012,Kumar2017,Lindquist2018,Ngyuen2018,Bommineni2018,Dshemuchadse2021}.
Bond orientational order parameters~\cite{Steinhardt1983} and related approaches~\cite{Honeycutt1987,Faken1994,Ackland2006,Stukowski2012,Malins2013,Martelli2018,Bozic2021} reliably detect the presence of global and local order and identify a few simple crystals, including face-centered cubic, hexagonal close-packed, and body-centered cubic~\cite{tenWolde1995,Auer2004,Lechner2008,Mickel2013,Eslami2018}.
A more general structure characterization requires local fingerprints~\cite{Keys2011,Tanaka2019} that are fed into machine learning algorithm~\cite{Geiger2013,Dietz2017,Reinhart2017,Boattini2019,Spellings2018,Boattini2018}.
Recent methods distinguish different complex crystals from another~\cite{Spellings2018,Boattini2018} but do not give crystallographic information.
In this contribution, we propose an algorithm to determine point group symmetry in particle simulation data.
The analysis starts with the spherical harmonics decomposition of the bond orientational order diagram, which can be rotated efficiently with the help of Wigner D-matrices.
We provide the necessary mathematical background and calculate Wigner D-matrices for selected transformations and point groups.

%%%%%%%%%%%%%%%%%%%%%%%%%%%%%%%%%%%%%%%%%%%%%%%%%%%%%%%%%%%%%%%%%%%%%%%%%%%%%%%%%%%%%%%%%%%%%%%%%%%%
%%%%%%%%%%%%%%%%%%%%%%%%%%%%%%%%%%%%%%%%%%%%%%%%%%%%%%%%%%%%%%%%%%%%%%%%%%%%%%%%%%%%%%%%%%%%%%%%%%%%
%%%%%%%%%%%%%%%%%%%%%%%%%%%%%%%%%%%%%%%%%%%%%%%%%%%%%%%%%%%%%%%%%%%%%%%%%%%%%%%%%%%%%%%%%%%%%%%%%%%%

\section{Mathematical Background}\label{Sec:mathematics}

Point group analysis combines mathematical methods from harmonic analysis, group theory, and statistics.
We briefly refresh definitions and notations in \cref{SSec:functionSphere,SSec:sphericalHarmonics,SSec:groupAction,SSec:WignerMatrices} and then expand them as necessary.
In particular, we introduce the Wigner D-matrix of a point group (\cref{SSec:WignerPG}), and propose the use of the covariance for point group analysis (\cref{SSec:CoVariance}).

\subsection{Functions on the sphere}\label{SSec:functionSphere}

Of central importance are functions defined on the surface of the unit sphere $S^2$.
We choose spherical coordinates $(\theta,\varphi)$.
The polar angle $\theta$ is measured from the $z$-axis and the azimuthal angle $\varphi$ is measured after orthogonal projection on the $xy$-plane from the $x$-axis.

Let $f(\theta,\varphi)$ and $f'(\theta,\varphi)$ be two such functions.
An inner product is defined by
\begin{equation}\label{Eq:innerproduct}
\langle f, f'\rangle=\frac{1}{4\pi}\int_{\theta=0}^\pi \int_{\varphi=0}^{2\pi} f(\theta,\varphi)\;f'(\theta,\varphi)^*\; \sin(\theta)\; d\varphi\; d\theta,
\end{equation}
where the superscript $^*$ denotes complex conjugation. 
A special case is the mean
\begin{equation*}
\langle f\rangle=\langle f, 1\rangle=\frac{1}{4\pi}\int_{\theta=0}^\pi \int_{\varphi=0}^{2\pi} f(\theta,\varphi)\; \sin(\theta)\; d\varphi\; d\theta.
\end{equation*}

The set of all functions defined on the surface of the unit sphere equipped with the addition of functions, scalar multiplication, and this inner product is a Hilbert space called $L^2(S^2)$.

\subsection{Spherical harmonics expansion}\label{SSec:sphericalHarmonics}

Particularly useful functions in $L^2(S^2)$ are the spherical harmonics
\begin{equation*}
Y_\ell^m(\theta,\varphi)~=~\sqrt{(2\ell+1) \frac{(\ell-m)!}{(\ell+m)!}}\; P_{\ell}^{m}(\cos \theta) \; \exp(im\varphi),
\end{equation*}
with the associated Legendre polynomials $P_{\ell}^{m}(\cos \theta)$.
Note that physicists often include an additional prefactor $(4\pi)^{-1/2}$.
We omit this prefactor by including it in the inner product because it would complicate many of the following expressions.

The normalization of spherical harmonics is chosen such that $\langle Y_\ell^m,Y_{\ell'}^{m'}\rangle=\delta_{\ell,\ell'}\;\delta_{m',m}$ with the Kronecker deltas $\delta_{\ell,\ell'}$ and $\delta_{m',m}$.
This means spherical harmonics are orthonormal.
It can be shown that the set of spherical harmonics $\{Y_\ell^m, 0\leq\ell<\infty, -\ell\leq m\leq\ell\}$ spans $L^2(S^2)$ and is therefore a basis~\cite{Axler2001}.
Spherical harmonics are the analogues of plane waves in flat space.

The spherical harmonics series expansion is the analogue of the Fourier series expansion.
It is defined as
\begin{equation}\label{Eq:expansion}
f(\theta,\varphi)=\sum_{\ell=0}^{\infty}\sum_{m=-\ell}^{\ell} Q_\ell^m Y_\ell^m(\theta,\varphi)
\end{equation}
with expansion coefficients
\begin{equation}\label{Eq:coefficients}
Q_\ell^m=\frac{1}{4\pi}\int_{\theta=0}^\pi \int_{\varphi=0}^{2\pi} f(\theta,\varphi)\;Y^m_\ell(\theta,\varphi)^*\; \sin(\theta)\; d\varphi\; d\theta
\end{equation}
or, in the notation of the inner product, $Q_\ell^m=\langle f, Y_\ell^m\rangle$.

The spherical harmonics expansion is invertible.
This means $f$ can be recovered from the set of all $\{Q_\ell^m\}_{\ell,m}$.

\subsection{Point group action}\label{SSec:groupAction}

Expansion coefficients depend on the choice of the coordinate system.
Let $g$ be a transformation from the old coordinates $(\theta,\varphi)$ to the new coordinates $(\theta',\varphi')$.
Such a transformation in Cartesian coordinates is an orthogonal $3\times3$ matrix.
It is a rotation, a reflection, or a combination of a rotation and a reflection.
The set of all coordinate transformations forms the orthogonal group $\OG$.

In the language of group theory we say that the action of $g$ on the function $f(\theta,\varphi)$ is an operation $L^2(S^2)\times\OG\rightarrow L^2(S^2)$ mapping $(f, g)$ to $f\cdot g$.
It is given by
\begin{equation}\label{Eq:groupaction}
g\cdot f(\theta,\varphi)= f(\theta',\varphi')=f'(\theta,\varphi),
\end{equation}
or short $g\cdot f = f'$, where $f'$ is the transformed function in new coordinates.
A special case is $e\cdot f=f$ for the identity transformation $e$.
The action of the product of two transformations $g$ and $g'$ is the composition $g\cdot (g' \cdot f) = (g g')\cdot f$.
In other words, the group action is compatible with the group operation.

A subgroup of $\OG$ is called a point group.
Let $G$ be a finite point group.
The symmetrization of $f$ under the action of $G$ is defined as the function
\begin{equation}\label{Eq:symmetrization}
f_G=\frac{1}{|G|}\sum_{g\in G} g\cdot f,
\end{equation}
where the order $|G|$ is the number of elements of $G$.
The sum averages over the action of all group elements.
Symmetrizations are invariant under the action of any of their own group elements, $g\cdot f_G = f_G$.

\subsection{Wigner matrices}\label{SSec:WignerMatrices}

The group action can be expressed with \cref{Eq:expansion} in the spherical harmonics basis as
\begin{equation}\label{Eq:basisExpansion}
g\cdot f
=\sum_{\ell=0}^{\infty}\sum_{m=-\ell}^{\ell} g\cdot (Q_\ell^m Y_\ell^m)
=\sum_{\ell=0}^{\infty}\sum_{m=-\ell}^{\ell} (g\cdot Q_\ell^m)\; Y_\ell^m.
\end{equation}
The orthogonal group now acts in this basis directly on the expansion coefficients.

It can be shown that the action of $g$ on the expansion coefficients mixes only expansion coefficients with different $m$ but not those with different $\ell$~\cite{Wigner1931}.
The group action in matrix form can therefore be written as
\begin{equation*}
g\cdot Q_\ell^{m'} = \sum_{m=-\ell}^{\ell}\WD_\ell^{m',m}(g) \; Q_\ell^{m}.
\end{equation*}
The matrix $\WD_\ell^{m',m}(g)$ is known as the Wigner D-matrix of the transformation $g$.
The set of all Wigner D-matrices is a matrix representation of $\OG$ with matrix product
\begin{equation*}
\WD_\ell^{m',m}(g g') = \sum_{m''=-\ell}^\ell \WD_\ell^{m',m''}(g) \; \WD_\ell^{m'',m}(g').
\end{equation*}

The Wigner D-matrix of a rotation is best calculated by relation to the real-valued Wigner d-matrix $\Wd_\ell^{m',m}(g)$ via
\begin{equation*}
\WD_\ell^{m',m}(g)=\exp(-im'\alpha)\;\Wd_\ell^{m',m}(\beta)\;\exp(-im\gamma),
\end{equation*}
where $(\alpha,\beta,\gamma)$ are Euler angles in the $z$--$y$--$z$ convention in a right-handed frame with right-hand screw rule and active interpretation.
In this convention, we first rotate by an angle $\alpha$ about the $z$-axis, then by an angle $\beta$ about the new $y$-axis, and finally by an angle $\gamma$ about the new $z$-axis.

Wigner matrix elements obey the relations
\begin{align}\label{Eq:Wignerrelations}
\Wd_\ell^{m',m}(\beta) &= (-1)^{m'+m} \Wd_\ell^{m,m'}(\beta)\nonumber\\
&=(-1)^{m'+m} \Wd_\ell^{-m',-m}(\beta)\nonumber\\
&=(-1)^{m'+m} \Wd_\ell^{m',m}(-\beta)\nonumber,\\
\WD_\ell^{m',m}(g) &= (-1)^{m'+m} \WD_\ell^{m,m'}(g)\\
&=(-1)^{m'+m} \WD_\ell^{-m',-m}(g)^*\nonumber\\
&=(-1)^{m'+m} \WD_\ell^{m',m}(g^{-1})^*.\nonumber
\end{align}

Wigner matrices can be calculated analytically via recursion and in a numerically fast and stable way~\cite{Gimbutas2009}.

\subsection{Wigner D-matrices of point groups}\label{SSec:WignerPG}

Wigner D-matrices can be averaged over group elements.
Let $G$ be a point group.
We define the Wigner D-matrix of $G$ as
\begin{equation}\label{Eq:averageWigner}
\WD_\ell^{m',m}(G)=\frac{1}{|G|}\sum_{g\in G}\WD_\ell^{m',m}(g).
\end{equation}

We extend Wigner D-matrices to semidirect products.
Semidirect products are useful because all point groups can be written as semidirect products of a few small point groups~\cite{Altmann1963a,Ezra1982}.
Let $G$ and $G'$ be two point groups such that the set $\{gg',g\in G,g'\in G'\}$ is again a point group.
Not all pairs of point groups fulfill this condition.
But if they do then we write
\begin{equation}
G\rtimes G'=\{gg',g\in G,g'\in G'\}.
\end{equation}
This set is called the semidirect product of $G$ and $G'$.

The behavior of Wigner D-matrices under the semidirect product is particularly simple.
We calculate
\begin{align}\label{Eq:semidirect}
\WD_\ell^{m',m}(G\rtimes G') &= \frac{1}{|G|}\frac{1}{|G'|}\sum_{g\in G}\sum_{g'\in G'}\WD_\ell^{m',m}(gg')\nonumber\\
&= \sum_{m''=-\ell}^\ell\WD_\ell^{m',m''}(G) \; \WD_\ell^{m'',m}(G').
\end{align}
This means Wigner D-matrices of a large point group are directly obtained from Wigner D-matrices of smaller point groups.

\subsection{Variance and covariance}\label{SSec:CoVariance}

The similarity of two functions in $f$ and $f'$ in $L^2(S^2)$ is measured with the help of the inner product $\langle f,f'\rangle$.
The covariance of $f$ and $f'$ is defined as
\begin{equation}
\CC(f,f')=\langle f - \langle f\rangle, f' - \langle f'\rangle\rangle.
\end{equation}
In the case of $f$ and $f'$ being equal, the covariance $K(f,f)$ is called the variance of $f$.

Of particular importance is the covariance of $f$ with its symmetrization $f_G$~\cite{Kazhdan2003, Kazhdan2004}.
The equations \cref{Eq:innerproduct}, \cref{Eq:groupaction}, \cref{Eq:symmetrization} give $\langle f, g\cdot f\rangle = \langle g^{-1}\cdot f, f\rangle$ and $\langle f \rangle = \langle f_G\rangle$ and $\langle f_G, f_G\rangle = \langle f, f_G\rangle$.
(Co)variances involving $f$ and $f_G$ therefore simplify to
\begin{align}
\CC(f_G,f_G)&=\CC(f,f_G),\nonumber\\
\CC(f,f_G)&=\langle f, f_G\rangle - |\langle f\rangle|^2,\\
\CC(f,f)&=\langle f, f\rangle - |\langle f\rangle|^2.\nonumber
\end{align}

Using the series expansion in \cref{Eq:basisExpansion} and the orthonormality of spherical harmonics, the (co)variances are expressed in matrix form as
\begin{align}\label{Eq:covariance}
\CC(f,f_G) &= \sum_{\ell=1}^{\infty}\sum_{m'=-\ell}^{\ell}\sum_{m=-\ell}^{\ell} Q_\ell^{m'}\;\WD_\ell^{m',m}(G)\; Q_\ell^{m*},\nonumber\\
\CC(f,f) &= \sum_{\ell=1}^{\infty}\sum_{m=-\ell}^{\ell} |Q_\ell^m|^2.
\end{align}
Note that both sums starts at $\ell=1$ and not at $\ell=0$.
The reason is the subtraction of $\langle f \rangle= Q_0^0$.

The covariance can be normalized by division with the square roots of the variances,
\begin{equation}
\nCC(f,f_G)=\frac{K(f,f_G)}{\sqrt{K(f,f)\;K(f_G,f_G)}} = \sqrt{\frac{K(f,f_G)}{K(f,f)}}.
\end{equation}
The normalized covariance is also known as the Pearson correlation coefficient.
It obeys $\nCC(f,f_G)=1$ if $f$ is invariant under the action of all $g\in G$ and $\nCC(f,f_G)<1$ otherwise.
We will use $\nCC(f,f_G)$ to analyze the presence of point group symmetry $G$ in $f$.

%%%%%%%%%%%%%%%%%%%%%%%%%%%%%%%%%%%%%%%%%%%%%%%%%%%%%%%%%%%%%%%%%%%%%%%%%%%%%%%%%%%%%%%%%%%%%%%%%%%%
%%%%%%%%%%%%%%%%%%%%%%%%%%%%%%%%%%%%%%%%%%%%%%%%%%%%%%%%%%%%%%%%%%%%%%%%%%%%%%%%%%%%%%%%%%%%%%%%%%%%
%%%%%%%%%%%%%%%%%%%%%%%%%%%%%%%%%%%%%%%%%%%%%%%%%%%%%%%%%%%%%%%%%%%%%%%%%%%%%%%%%%%%%%%%%%%%%%%%%%%%

\section{Examples of Wigner D-matrices}

We calculate in this section Wigner D-matrices for selected transformations and point groups.
The main results are summarized in \cref{Table:Wigner}.

\begin{table}
\begin{center}
\setlength{\tabcolsep}{3mm}
\begin{tabular}{c l l}
\hline\hline
\noalign{\vskip 1mm}
$*$ & $\WD_\ell^{m',m}(*)$ & Description\\
\noalign{\vskip 1mm}
\hline
\noalign{\vskip 1mm}
$e$ & $\delta_{m',m}$ &identity\\
$i$ & $\delta_{m',m}(-1)^\ell$ & inversion\\
$\sigma_{yz}$ & $\delta_{m',-m}$ & reflection in $yz$-plane\\
$\sigma_{xz}$ & $\delta_{m',-m}(-1)^{m}$ & reflection in $xz$-plane\\
$\sigma_{xy}$ & $\delta_{m',m}(-1)^{m+\ell}$ & reflection in $xy$-plane\\
$2_x$ & $\delta_{m',-m} (-1)^\ell$ & rotation about $x$-axis\\
$2_y$ & $ \delta_{m',-m} (-1)^{m+\ell}$ & rotation about $y$-axis\\
$n_z$ & $\delta_{m',m}\; \exp(-2\pi i \, m / n)$ & rotation about $z$-axis\\
\noalign{\vskip 1mm}
\hline
\noalign{\vskip 1mm}
$C_i$ & \multicolumn{2}{l}{$\delta_{m',m}\;\delta_{\ell\bmod 2,0}$}\\
$C_n$ & \multicolumn{2}{l}{$\delta_{m',m}\;\delta_{m \bmod n,0}$}\\
$C_\infty$ & \multicolumn{2}{l}{$\delta_{m',m}\;\delta_{m,0}$}\\
$D_n$ & \multicolumn{2}{l}{$\frac{1}{2}(\delta_{m',m}+\delta_{m',-m} (-1)^\ell)\delta_{m \bmod n,0}$}\\
$T$ & \multicolumn{2}{l}{
$\WD_0^{0,0}\!=\!1$,\;
$\WD_3^{2,2}\!=\!\frac{1}{2}$,\;
$\WD_4^{0,0}\!=\!\frac{7}{12}$,\;
$\WD_4^{4,0}\!=\!\frac{\sqrt{70}}{24}$,}\\
& \multicolumn{2}{l}{
$\WD_4^{4,4}\!=\!\frac{5}{24}$,\;
$\WD_6^{0,0}\!=\!\frac{1}{8}$,\;
$\WD_6^{2,2}\!=\!\frac{11}{32}$,\;
$\WD_6^{4,0}\!=\!-\frac{\sqrt{14}}{16}$,}\\
& \multicolumn{2}{l}{
$\WD_6^{4,4}\!=\!\frac{7}{16}$,\;
$\WD_6^{6,2}\!=\!-\frac{\sqrt{55}}{32}$,\;
$\WD_6^{6,6}\!=\!\frac{5}{32}$}\\
$O$ & \multicolumn{2}{l}{
$\WD_0^{0,0}\!=\!1$,
$\WD_4^{0,0}\!=\!\frac{7}{12}$,\;
$\WD_4^{4,0}\!=\!\frac{\sqrt{70}}{24}$,\;
$\WD_4^{4,4}\!=\!\frac{5}{24}$,}\\
& \multicolumn{2}{l}{
$\WD_6^{0,0}\!=\!\frac{1}{8}$,\;
$\WD_6^{4,0}\!=\!-\frac{\sqrt{14}}{16}$,\;
$\WD_6^{4,4}\!=\!\frac{7}{16}$}\\
$I$ & \multicolumn{2}{l}{
$\WD_0^{0,0}\!=\!1$,\;
$\WD_6^{0,0}\!=\!\frac{11}{25}$,\;
$\WD_6^{5,0}\!=\!\frac{\sqrt{77}}{25}$,\;
$\WD_6^{5,5}\!=\!\frac{7}{25}$}\\
\noalign{\vskip 1mm}
\hline\hline
\end{tabular}
\caption{
Wigner D-matrix coefficients $\WD_\ell^{m',m}$ for selected transformations (top part of the table) and point groups (bottom part).
The Schoenflies notation is used to denote point groups and the orientation of the coordinate system explained in the text.
For transformations, a description of the symmetry is provided.
For the polyhedral groups $T$, $O$, $I$, only coefficients with $m',m>0$, $m'>m$ and $\ell\leq6$ are listed.
Coefficients with other values of $m$ and $m'$ can be calculated using the relations \cref{Eq:Wignerrelations2}.
}
\label{Table:Wigner}
\end{center}
\end{table}

\subsection{Reflection symmetry}\label{SubSec:Reflection}

The simplest Wigner D-matrix is that of the identity transformation,
\begin{equation}
\WD_\ell^{m',m}(e) = \delta_{m',m}.
\end{equation}

The reflection in the $yz$-plane, denoted by $\sigma_{yz}$, maps the direction $(\theta,\varphi)$ on $(\theta,\pi-\varphi)$.
Likewise, the reflection in the $xz$-plane, denoted by $\sigma_{xz}$, maps $(\theta,\varphi)$ on $(\theta,-\varphi)$, and the reflection in the $xy$-plane, denoted by $\sigma_{xy}$, maps $(\theta,\varphi)$ on $(\pi-\theta,\varphi)$.
From the definition of spherical harmonics we obtain
\begin{align}
\WD_\ell^{m',m}(\sigma_{yz}) &= \delta_{m',-m}\nonumber\\
\WD_\ell^{m',m}(\sigma_{xz}) &= \delta_{m',-m}(-1)^{m}\\
\WD_\ell^{m',m}(\sigma_{xy}) &= \delta_{m',m}(-1)^{m+\ell}\nonumber.
\end{align}
The composition of all three reflections is the inversion $i=\sigma_{yz}\sigma_{xz}\sigma_{xy}$ with
\begin{equation}
\WD_\ell^{m',m}(i) = \delta_{m',m}(-1)^\ell.
\end{equation}

As an example, we calculate the Wigner D-matrix of the inversion group $C_i$, which is generated by the inversion $i$.
We write in Schoenflies notation $C_i=\langle i\rangle = \{e,i\}$.
Note that the use of angle brackets $\langle\cdot\rangle$ for group generators differs from their use for inner products in \cref{Eq:innerproduct}.
Both notations are standard in their respective fields.
Averaging over contributions from its two elements gives
\begin{equation}
\WD_\ell^{m',m}(C_i) = \delta_{m',m}\frac{1 + (-1)^\ell}{2} = \delta_{m',m}\;\delta_{\ell\bmod 2,0}.
\end{equation}
The second Kronecker delta uses the modulo operation, which guarantees that only matrix elements with even $\ell$ are non-zero.

\subsection{Rotational symmetry}

Rotations about the $z$-axis are particularly simple because the Euler angle $\beta$ is zero.
The rotation $n_z$ about the $z$-axis by the angle $2\pi / n$ corresponds to the Wigner D-matrix
\begin{equation}
\WD_\ell^{m',m}(n_z)=\delta_{m',m}\; \exp(-2\pi i \, m / n).
\end{equation}
We calculate Wigner D-matrices for the cyclic group $C_n=\langle n_z \rangle=\{e,n_z,n_z^2,\ldots,n_z^{n-1}\}$ by averaging over contributions from all group elements.
Most sums are zero with the exception of the cases where $m$ an integer multiple of $n$,
\begin{equation}
\WD_\ell^{m',m}(C_n) = \delta_{m',m}\;\delta_{m \bmod n,0}.
\end{equation}

The Wigner D-matrix of continuous axial symmetry is obtained by taking $n$ to $\infty$ as
\begin{equation}
\WD_\ell^{m',m}(C_\infty) = \lim_{n\rightarrow\infty}\WD_\ell^{m',m}(C_n) = \delta_{m',m}\;\delta_{m,0}.
\end{equation}

Rotations about other axis have more complicated coefficients that are cumbersome to use~\cite{Altmann1957,Altmann1963}.
Exceptions are the two-fold rotation about the $x$-axis, $2_x=\sigma_{yz}i$, and the two-fold rotation about the $y$-axis, $2_y=\sigma_{xz}i$. 
The corresponding Wigner D-matrices are
\begin{align}\label{Eq:Wigner2-fold}
\WD_\ell^{m',m}(2_x) &= \delta_{m',-m} (-1)^\ell,\nonumber\\
\WD_\ell^{m',m}(2_y) &= \delta_{m',-m} (-1)^{m+\ell}.
\end{align}

The dihedral group $D_n=\langle n_z,2_x\rangle$ has $2n$ elements and is the semidirect product $D_n=C_n\rtimes \langle 2_x\rangle$.
From \cref{Eq:semidirect} we obtain
\begin{equation}
\WD_\ell^{m',m}(D_n) = \frac{\delta_{m',m}+\delta_{m',-m} (-1)^\ell}{2}\delta_{m \bmod n,0}.
\end{equation}
Another common semidirect product is the semidirect product with the group $C_i$.
This semidirect product includes a factor $\delta_{\ell\bmod 2,0}$.
For example, for the point group $C_{4h}=C_4\rtimes C_i$ we have
\begin{equation}
\WD_\ell^{m',m}(C_{4h}) = \delta_{m',m}\;\delta_{m \bmod 4,0}\;\delta_{\ell\bmod 2,0}.
\end{equation}
Wigner D-matrices of all axial point groups can be obtained via appropriate semidirect products.

\subsection{Polyhedral symmetry}

Point groups are said to be of polyhedral symmetry if they have more than one rotation axis of order greater than two.
The tetrahedral group $T$, the octahedral group $O$, and the icosahedral group $I$ are given by
\begin{align}\label{Eq:polyGroups}
T=\langle 2_z, 3_{1,1,1}\rangle\quad&\text{with}\quad |T|=12,\nonumber\\
O=\langle 4_z, 3_{1,1,1}\rangle\quad&\text{with}\quad |O|=24,\\
I=\langle 5_z, 2_{1,0,\tau}\rangle\quad&\text{with}\quad |I|=60,\nonumber
\end{align}
where $3_{1,1,1}$ is a three-fold rotation about the axis $(1,1,1)$ and $2_{1,0,\tau}$ is a two-fold rotation about the axis $(1,0,\tau)$ with the golden mean $\tau=(1+\sqrt{5})/2$.

The Wigner D-matrices of the polyhedral groups do not have an expression we can write down.
But the choice of coordinates in \cref{Eq:polyGroups} guarantees the simplest form we could find.
It has the special properties that each polyhedral group contains $2_y$.
This is helpful because from \cref{Eq:Wignerrelations} and \cref{Eq:Wigner2-fold} it follows that any Wigner D-matrix of a point group that contains $2_y$ is real-valued and obeys the relations
\begin{align}\label{Eq:Wignerrelations2}
\WD_\ell^{m',m}(g) &= (-1)^{m'+m} \WD_\ell^{m,m'}(g),\nonumber\\
&= (-1)^{m'+\ell} \WD_\ell^{-m',m}(g),\nonumber\\
&= (-1)^{m+\ell} \WD_\ell^{m',-m}(g).
\end{align}

Coefficients of Wigner D-matrices of the three polyhedral groups for $\ell\leq 6$ are listed in \cref{Table:Wigner}.

\subsection{Wigner D-matrix traces}

\begin{table}
\begin{center}
\setlength{\tabcolsep}{2mm}
\begin{tabular}{c l l}
\hline\hline
\noalign{\vskip 1mm}
Group & \multicolumn{2}{l}{Data for Wigner D-matrix traces $\WD_\ell(G)$}\\
\noalign{\vskip 1mm}
\hline
\noalign{\vskip 1mm}
$C_i$ & \multicolumn{2}{l}{$\WD_\ell(C_i)=\lfloor \ell / 2\rfloor + 1$}\\
$C_n$ & \multicolumn{2}{l}{$\WD_\ell(C_n)=2\lfloor \ell / n\rfloor + 1$}\\
$D_n$ & \multicolumn{2}{l}{$\WD_\ell(D_n)=\lfloor \ell / n\rfloor + (-1)^\ell$}\\
$T$ & $r=6$ & $b=100110$\\
$O$ & $r=12$ & $b=100010101110$\\
$I$ & $r=30$ & $b=100000100010100110101110111110$\\
\noalign{\vskip 1mm}
\hline\hline
\end{tabular}
\caption{
Data for Wigner D-matrix traces for selected point groups.
Formulas for groups with rotational symmetry are directly provided.
Traces for polyhedral groups are calculated using $\WD_\ell(G)=\lfloor \ell / r\rfloor + b[\ell\bmod r]$ with repeat length $r$, the floor function $\lfloor\;\rfloor$, and a number (0 or 1) selected by index from the block string $b$.
For example, $b[5]$ returns the fifth number of the block $b$.
}
\label{Table:WignerTraces}
\end{center}
\end{table}

Wigner D-matrices depend on the choice of coordinate system.
Functionals that are invariant under rotation are the traces
\begin{align*}
\WD_\ell(G) = \sum_{m=-\ell}^\ell \WD_\ell^{m,m}(G).
\end{align*}
We call $\WD_\ell(G)$ the Wigner D-matrix trace of the point group $G$ for the given $\ell$.
Calculating the traces is helpful because we know that if $\WD_\ell(G)=0$ then all coefficients $\{Q_\ell^m\}_m$ are zero.
We then say these coefficients are zero due to systematic extinction.

Wigner D-matrix traces are listed in \cref{Table:WignerTraces}.
Traces for the point groups $C_i$, $C_n$, and $D_n$ are calculate directly from \cref{Table:Wigner}.
Traces for polyhedral point groups are determined algorithmically.
We observe that the traces form blocks.
The pattern repeats in each block except that the coefficients increase by one from a block to the next.
For example,
\begin{align*}
(\WD_\ell(D_3))_\ell &= 1,0,1,\;\;1,2,1,\;\;3,2,3,\;\;3,4,3,\;\;\ldots\nonumber\\
(\WD_\ell(T))_\ell &= 1,0,0,1,1,0,\;\;2,1,1,2,2,1,\;\;\ldots
\end{align*}
with gaps indicating the end of blocks.
The first non-zero Wigner D-matrix coefficients for $\ell>0$ is found at $\ell=3$ for tetrahedral symmetry, at $\ell=4$ for octahedral symmetry, and at $\ell=6$ for icosahedral symmetry.

%%%%%%%%%%%%%%%%%%%%%%%%%%%%%%%%%%%%%%%%%%%%%%%%%%%%%%%%%%%%%%%%%%%%%%%%%%%%%%%%%%%%%%%%%%%%%%%%%%%%
%%%%%%%%%%%%%%%%%%%%%%%%%%%%%%%%%%%%%%%%%%%%%%%%%%%%%%%%%%%%%%%%%%%%%%%%%%%%%%%%%%%%%%%%%%%%%%%%%%%%
%%%%%%%%%%%%%%%%%%%%%%%%%%%%%%%%%%%%%%%%%%%%%%%%%%%%%%%%%%%%%%%%%%%%%%%%%%%%%%%%%%%%%%%%%%%%%%%%%%%%

\section{Bond orientational order diagrams}

Information about orientational order in particle simulation data is contained in the directions of neighbor bonds.
Bond orientational order diagrams are functions defined on the surface of the sphere.
We first introduce bond orientational order diagrams and then apply the theory developed in \cref{Sec:mathematics} to them.

\subsection{Definition}

Let $\{(x_i,y_i,z_i),i=1,\ldots,N\}$ be a set of $N$ particle coordinates in a system.
The direction of the bond connecting particle $i$ to particle $j\neq i$ is given in spherical coordinates by $(\theta_{ij},\varphi_{ij})$ with
\begin{align*}
\theta_{ij} &= \arccos\frac{z_j-z_i}{\sqrt{(x_j-x_i)^2+(y_j-y_i)^2+(z_j-z_i)^2}},\nonumber\\
\varphi_{ij} &= \arctantwo(y_j-y_i,x_j-x_i),
\end{align*}
where $\arctantwo$ is the two-argument arctangent function popular in programming languages that takes into account the correct quadrant of $(x,y)$.
If the system employs periodic boundary conditions then the bond direction is chosen as the direction of the bond connecting particle $i$ with the closest image of particle $j$.

The bond orientational order diagram is then the function in $L^2(S^2)$ defined by
\begin{equation*}
b(\theta,\varphi)=\frac{4\pi}{\sin(\theta)}\sum_{i,j\neq i}\frac{w_{ij}}{w}\;\delta(\theta-\theta_{ij})\;\delta(\varphi-\varphi_{ij}),
\end{equation*}
where the sum runs over all particle pairs and $\delta$ is the Dirac delta function.
The $w_{ij}$ are called bond weights.
Bond weights select bonds (if non-zero) and weight the contribution of each bond to the bond orientational order diagram.
The normalization by division by the total bond weight $w=\sum_{i,j\neq i} w_{ij}$ guarantees convergence in the thermodynamic limit $N\rightarrow\infty$.

We perform a spherical harmonics expansion of the bond orientational order diagram.
The expansion coefficients \cref{Eq:coefficients} now have the expression
\begin{equation}\label{Eq:expansionCoefficients}
Q_\ell^m=\sum_{i,j\neq i}\frac{w_{ij}}{w}Y_\ell^{m}(\theta_{ij},\varphi_{ij})^*.
\end{equation}
Because the expansion coefficients are not invariant under rotation, it is customary in physics and materials simulations~\cite{Steinhardt1983,tenWolde1995,Auer2004,Lechner2008,Mickel2013} to calculate the rotationally invariant combinations
\begin{equation}
Q_\ell=\sqrt{\frac{1}{2\ell+1}\sum_{m=-\ell}^{\ell}|Q_\ell^m|^2},
\end{equation}
which are obtained by averaging over contributions for all $m$ with fixed $\ell$.
The $Q_\ell$ are known as the global bond orientational order parameters.
They are essentially quadratic forms in the expansion coefficients, just like the (co)variances in \cref{Eq:covariance}.

Bond orientational order parameters are easy to use because they are rotationally invariant.
Their application in condensed matter physics follows empirical rules.
For example, Steinhardt et al.~\cite{Steinhardt1983} proposed $Q_4$ for systems with cubic symmetry and $Q_6$ for systems with icosahedral symmetry.
This proposition is based on the observation that these numbers are ``the first non-zero average'', a fact for which the authors cite a private communication with N.\ David Mermin.
Indeed, our calculations of Wigner D-matrix traces in \cref{Table:WignerTraces} confirms that $\WD_4(O)$ and $\WD_6(I)$ are the first non-zero traces.
Beyond the use of bond orientational order parameters to detect order, they are not helpful for analyzing crystallographic order.
The reason is the loss of information by spherical averaging.
In the following, we propose new order parameters that utilize the complete information contained in the expansion coefficients $\{Q_\ell^m\}_{\ell,m}$.

\subsection{Bond weighting techniques}

The precise way to choose bond weights is not important for the methods developed in this work.
The bond weighting technique most commonly used in the literature uses a cut-off~\cite{Steinhardt1983}, which means $w_{ij}=H(r_\text{c}-\|\vec{x}_i-\vec{x}_j\|)$ with cut-off distance $r_\text{c}$.
$H$ is the Heaviside step function.
The total bond weight in cut-off weighting is $w=N_\text{B}$ with $N_\text{B}$ being the number of bonds in the system that have bond length smaller than $r_\text{c}$.

Another suitable technique for selecting bond weights is Voronoi cell weighting~\cite{Mickel2013}.
Voronoi cell weighting sets $w_{ij}$ equal to the contact area of the Voronoi cell of particles $i$ with the Voronoi cell of particle $j$.
Bond weights are automatically zero if the corresponding particles are far enough apart because their Voronoi cells cannot be in face-to-face contact.
Voronoi cell weighting is more robust under small particle motions but requires more efforts to implement and evaluate.

Both bond weighting techniques, cut-off weighting and Voronoi cell weighting, have in common that they generate symmetric bond weights, $w_{ij}=w_{ji}$.
A bond orientational order diagram with symmetric bond weights is by definition inversion symmetric.
Its point symmetry can then be written as the semidirect product with the group $C_i$.
As a consequence, as shown in \cref{SubSec:Reflection}, the Wigner D-matrix contains a factor $\delta_{\ell\bmod 2,0}$.
This means the coefficients $\WD_\ell^{m',m}$ and $Q_\ell^m$ of any bond orientational order diagram with symmetric bond weights are zero unless $\ell$ is an even number.

\subsection{Covariance of the fluid}

Bond directions of particles sufficiently far enough separated are uncorrelated in fluids.
This means the bond orientational order diagram of a fluid, denoted by $\bfl$, becomes constant in the thermodynamic limit, $\lim_{N\rightarrow\infty}\bfl(\theta,\varphi) = 1$.
All but one expansion coefficient are zero in the fluid,
\begin{equation*}
Q_\ell^m = \langle \bfl,Y_\ell^m\rangle = \langle \bfl\rangle \langle Y_\ell^m\rangle = \delta_{m,0}\;\delta_{\ell,0}\;Y_0^0.
\end{equation*}
Fluids do not have bond orientational order for large $N$.
The same is not true for the covariance for finite $N$.

We calculate quadratic terms that appear in the covariance.
To first order in $N^{-1}$ the products of expansion coefficients are given by
\begin{align}\label{Eq:omegaDefinition}
Q_\ell^{m'} Q_\ell^{m*} &= \langle \bfl,Y_\ell^{m'}\rangle \langle Y_\ell^m,\bfl\rangle = \langle\bfl,\bfl\rangle\;\delta_{m',m} \nonumber\\
&= \sum_{i,j\neq i}\frac{w_{ij}^2}{w^2}\;\delta_{m',m}= \omega\;\delta_{m',m}
\end{align}
with prefactor $\omega=\sum_{i,j\neq i}w_{ij}^2/w^2$.
In the case of cut-off weighting, the prefactor is $\omega=1/N_\text{B}$.

Substitution into \cref{Eq:covariance} determines (co)variances of $\bfl$ with its symmetrization $\bfl_G$,
\begin{align}\label{Eq:covarianceFluid}
\CC(\bfl,\bfl_G) &= \omega\sum_{\ell=1}^{\ell_\text{max}}\sum_{m=-\ell}^{\ell} \WD_\ell^{m,m}(G),\nonumber\\
\CC(\bfl,\bfl) &= \omega\sum_{\ell=1}^{\ell_\text{max}}\sum_{m=-\ell}^{\ell} 1 = \omega\ell_\text{max}(\ell_\text{max}+2).
\end{align}
Here we limit the sum over $\ell$ to below $\ell_\text{max}$ because the sums do not converge.

Despite the disordered nature of the fluid, the normalized covariance is not zero.
Examples for a large enough expansion limit $\ell_\text{max}\gg 1$ are
\begin{align*}
\nCC(\bfl,\bfl_{C_n}) &= n^{-1/2},\nonumber\\
\nCC(\bfl,\bfl_{D_n}) &= (4n)^{-1/2}.
\end{align*}
Covariances for polyhedral groups are evaluated only numerically.

\subsection{Order parameters}

\begin{table}
\begin{center}
\setlength{\tabcolsep}{1mm}
\begin{tabular}{l l l}
\hline\hline
\noalign{\vskip 1mm}
\emph{Input:} & bond directions $(\theta_{ij},\varphi_{ij}\}_{ij}$, weights $\{w_{ij}\}_{ij}$,\\
& point group $G$, expansion limit $\ell_\text{max}$\\
\noalign{\vskip 3mm}
\emph{Step 1:} & \(\displaystyle Q_\ell^m=\sum_{i,j\neq i}\frac{w_{ij}}{w}Y_\ell^m(\theta_{ij},\varphi_{ij})^* \) & \cref{Eq:expansionCoefficients}\\
& \(\displaystyle \WD_\ell^{m',m}(G)=\frac{1}{|G|}\sum_{g\in G}\WD_\ell^{m',m}(g) \) & \cref{Eq:averageWigner}\\
& \(\displaystyle \omega=\sum_{i,j\neq i}\frac{w_{ij}^2}{w^2} \) & \cref{{Eq:omegaDefinition}}\\
\noalign{\vskip 3mm}
\emph{Step 2:} & \(\displaystyle QDQ=\sum_{\ell=1}^{\ell_\text{max}}\sum_{m'=-\ell}^{\ell}\sum_{m=-\ell}^{\ell} Q_\ell^{m'}\;\WD_\ell^{m',m}(G)\; Q_\ell^{m*} \) & \cref{Eq:covariance}\\
& \(\displaystyle QEQ=\sum_{\ell=1}^{\ell_\text{max}}\sum_{m=-\ell}^{\ell} |Q_\ell^{m}|^2 \) & \cref{Eq:covariance}\\
& \(\displaystyle EDE=\omega\sum_{\ell=1}^{\ell_\text{max}}\sum_{m=-\ell}^{\ell} \WD_\ell^{m,m}(G) \) & \cref{Eq:covarianceFluid}\\
& \(\displaystyle EEE=\omega\sum_{\ell=1}^{\ell_\text{max}}\sum_{m=-\ell}^{\ell} 1= \omega\ell_\text{max}(\ell_\text{max}+2) \) & \cref{Eq:covarianceFluid}\\
\noalign{\vskip 3mm}
\emph{Step 3:} & \(\displaystyle S = \frac{QEQ}{EEE} - 1 \) & \cref{Eq:totalOrderParameter}\\
\noalign{\vskip 1mm}
& \(\displaystyle S_G = \frac{\frac{QDQ}{QEQ} - \frac{EDE}{EEE}} {1 - \frac{EDE}{EEE}} \) & \cref{Eq:orderParameter}\\
\noalign{\vskip 1mm}
\emph{Output:} & order parameters $S$ and $S_G$,\\
& $S>0.5$: system starts to order\\
& $S_G>0.75$: system has point group $G$\\
\noalign{\vskip 1mm}
\hline\hline
\end{tabular}
\caption{
Algorithm for calculating the total order parameter $S$ and the point group symmetry order parameter $S_G$ in the bond orientational order diagram.
Step 2 introduces a shortened notation for the matrix sums representing (co)variances.
References to the equation numbers in the text that introduce each expression are provided in the right-hand column.}
\label{Table:orderParameter}
\end{center}
\end{table}

We now have all tools at hand to introduce new order parameters that utilize the complete information contained in the expansion coefficients.
Let $b$ be a bond orientational order diagram and $G$ a point group.
We define the point group symmetry order parameter as
\begin{equation}\label{Eq:orderParameter}
S_G(b) = \frac{\nCC(b,b_G)^2 - \nCC(\bfl,\bfl_G)^2}{1 - \nCC(\bfl,\bfl_G)^2}.
\end{equation}
The definition is chosen such that the order parameter is zero for the reference state and normalized for the target symmetry.
In other words it is zero in the fluid phase, $S_G(\bfl)=0$, and one if $b$ has full point group symmetry, $S_G(b_G)=1$.

Because the expansion coefficients $Q_\ell^m$ for $\ell>0$ contain all the information about orientational order, we define the total order parameter as the ratio of the covariance of the bond orientational order parameter of the system of interest to the covariance of the bond orientational order parameter of the fluid,
\begin{equation}\label{Eq:totalOrderParameter}
S(b) = \frac{\CC(b,b)}{\CC(\bfl,\bfl)}-1.
\end{equation}
The total order parameter analyzes the presence of any form of orientational order independent of a specific point group.
It is zeroed, $S(\bfl)=0$, but not normalized.

The algorithm to calculate all order parameters introduced in this work is listed in \cref{Table:orderParameter}.
It takes as input information about the bonds and the point group.
The outputs are the total order parameter $S$ and the point group symmetry parameters $S_G$.
Coefficients of the spherical harmonics expansion are considered in the algorithm only for $\ell\leq\ell_\text{max}$.
This means we omit high-frequency modulations of the bond orientational order diagram.

With the help of the order parameters we can detect the presence of order in the system using fixed threshold values.
We observe empirically that $S>0.5$ is a good threshold for the initial development of order somewhere in the system, for example in the form of small nuclei.
We also find that $S_G>0.75$ is a good threshold to determine whether the bond orientational order diagram exhibits point group symmetry $G$.

%%%%%%%%%%%%%%%%%%%%%%%%%%%%%%%%%%%%%%%%%%%%%%%%%%%%%%%%%%%%%%%%%%%%%%%%%%%%%%%%%%%%%%%%%%%%%%%%%%%%
%%%%%%%%%%%%%%%%%%%%%%%%%%%%%%%%%%%%%%%%%%%%%%%%%%%%%%%%%%%%%%%%%%%%%%%%%%%%%%%%%%%%%%%%%%%%%%%%%%%%
%%%%%%%%%%%%%%%%%%%%%%%%%%%%%%%%%%%%%%%%%%%%%%%%%%%%%%%%%%%%%%%%%%%%%%%%%%%%%%%%%%%%%%%%%%%%%%%%%%%%

\section{Discussion and Conclusion}

Just like its experimental counterpart, crystal structure analysis from particle simulation data will likely be performed in four steps that gradually extract different aspects of structural order.
In the first step, high-symmetry directions must be identified and the crystal oriented.
In the second step, rotational symmetry in form of a point group is determined.
Both of these steps are solved by the algorithms presented in this work.
All that is needed in practice is the recipe in~\cref{Table:orderParameter}.
In the third step, translational symmetry and the space group must be obtained.
Finally, in the fourth step, extracting a unit cell decoration completes crystal structure analysis.
The last two steps are open problems for future work.

Starting crystal structure analysis with the point group makes sense because orientational order typically develops first and is more pronounced whereas translational order is easily disturbed by defects such as disclinations and stacking faults.
Besides crystal structure analysis, the algorithms presented here are also applicable for the analysis of orientational order in finite clusters~\cite{Wales1997, Wang2018} and for the determination of molecular symmetry~\cite{Ivanov1999, Largent2012}.

\begin{acknowledgments}
	This work has been funded by Deutsche Forschungsgemeinschaft through Project EN 905/4-1,
	Support by the Central Institute for Scientific Computing), the Interdisciplinary Center for Functional Particle Systems, and computational resources and support provided by the Erlangen Regional Computing Center (RRZE) are gratefully acknowledged.
\end{acknowledgments}

\section*{Bibliography}
%aipnum4-2.bst 2019-01-14 (MD) hand-edited version of apsrev4-1.bst

\end{document}